\documentclass[fleqn,11pt]{SelfArx} 
\usepackage[english]{babel} 
\usepackage{epsfig}
\usepackage{float}
\usepackage{lipsum} 
\usepackage{fancybox,framed}
\usepackage{url}
\usepackage[varg]{newtxmath}
\usepackage[font=small,labelfont=bf,
   justification=justified,
   format=plain]{caption} 

\def\hitomi{{\textit{Hitomi}}}
\def\athena{{\textit{Athena}}}
\def\xrism{{\textit{XRISM}}}

\def\chandra{{\textit{Chandra}}}
\def\xmm{{\textit{XMM-Newton}}}
\def\nustar{{\textit{NuSTAR}}}

\def\ltsim{\mathrel{\hbox{\rlap{\hbox{\lower3pt\hbox{$\sim$}}}\hbox{$<$}}}}
\def\gtsim{\mathrel{\hbox{\rlap{\hbox{\lower3pt\hbox{$\sim$}}}\hbox{$>$}}}}

\definecolor{color1}{rgb}{0.0,0.0,.0} 
\definecolor{color2}{rgb}{1.0, 1.0, 1.0} 
\definecolor{color3}{rgb}{1.0,1.0,1.0} 
\definecolor{color4}{rgb}{0.0,0.0,0.0} 
\definecolor{colorbox}{rgb}{0.925, 0.956, 0.992}
\definecolor{colorheader}{rgb}{0.33,0.41,0.87} 
\definecolor{mygreen}{rgb}{0.13,0.55,0.13} 
\definecolor{mypurple}{rgb}{0.33,0.23,0.83} 

\usepackage{hyperref} 
\hypersetup{hidelinks,colorlinks,breaklinks=true,urlcolor=color2,citecolor=color1,linkcolor=color1,bookmarksopen=false,pdftitle={Title},pdfauthor={Author}}

\JournalInfo{Nat. As. Review} 
\Archive{} 

\PaperTitle{Accretion physics at high X-ray spectral resolution:} 
\PaperSubTitle{New frontiers and game-changing science} 

\Authors{{\em P.\,Gandhi,\textsuperscript{1} T.\,Kawamuro,\textsuperscript{2} M.\,D\'{i}az Trigo,\textsuperscript{3} J.A.\,Paice,\textsuperscript{1, 4} P.G.\,Boorman,\textsuperscript{5, 1} M.\,Cappi,\textsuperscript{6} C.\,Done,\textsuperscript{7} A.C.\,Fabian,\textsuperscript{8} K.\,Fukumura,\textsuperscript{9} J.A.\,Garc\'{i}a,\textsuperscript{10} C.L.\,Greenwell,\textsuperscript{1, 3}
M.\,Guainazzi,\textsuperscript{11}
K.\,Makishima,\textsuperscript{12, 13, 14} M.S.\,Tashiro,\textsuperscript{15} R.\,Tomaru,\textsuperscript{7} 
F.\,Tombesi,\textsuperscript{16, 17, 18, 19} Y.\,Ueda\textsuperscript{20}
}}
\affiliation{\textsuperscript{1}\textit{School of Physics \& Astronomy, University of Southampton, Highfield, Southampton SO17\,1BJ, UK}} 
\affiliation{\textsuperscript{2}\textit{Nu\'{c}leo de Astronom\'{i}a de la Facultad de Ingenier\'{i}a, Universidad Diego Portales, Av. Ej\'{e}rcito Libertador 441, Santiago, Chile}} 
\affiliation{\textsuperscript{3}\textit{European Southern Observatory, Karl-Schwarzschild-Strasse 2, 85748 Garching bei M\:{u}nchen, Germany}} 
\affiliation{\textsuperscript{4}\textit{Inter-University Centre for Astronomy \& Astrophysics, Pune University, Ganeshkhind, Pune 411007, India}}
\affiliation{\textsuperscript{5}\textit{Astronomical Institute of the Czech Academy of Sciences, Bo\v{c}n\'{i} II 1401, CZ-14100 Prague, Czechia}}
\affiliation{\textsuperscript{6}\textit{INAF -- OAS Bologna, Astrophysics and Space Science Observatory,  Via Gobetti 101, Bologna, Italy}}
\affiliation{\textsuperscript{7}\textit{Centre for Extragalactic Astronomy, Department of Physics, University of Durham, South Road, Durham, DH1\,3LE, UK}}
\affiliation{\textsuperscript{8}\textit{Institute of Astronomy, University of Cambridge, Madingley Road, Cambridge CB3\,0HA, UK}}
\affiliation{\textsuperscript{9}\textit{Department of Physics and Astronomy, James Madison University, Harrisonburg, VA 22807, USA)}}
\affiliation{\textsuperscript{10}\textit{Cahill Center for Astronomy and Astrophysics, California Institute of Technology, Pasadena, CA 91125, USA}}
\affiliation{\textsuperscript{11}\textit{ESA European Space Research and Technology Centre (ESTEC), Keplerlaan 1, NL-2201 AZ Noordwijk, the Netherlands}}
\affiliation{\textsuperscript{12}\textit{Kavli Institute for the Physics and Mathematics of the Universe, 5-1-5 Kashiwa-no-ha, Kashiwa, Chiba 277-8683, Japan}}
\affiliation{\textsuperscript{13}\textit{Department of Physics, The University of Tokyo, 7-3-1 Hongo, Bunkyo-ku, Tokyo 113-0033, Japan}} \affiliation{\textsuperscript{14}\textit{The Institute of Physical and Chemical Research (RIKEN), 2-1 Hirosawa, Wako, Saitama 351-0198, Japan}}
\affiliation{\textsuperscript{15}\textit{Department of Physics, Saitama University, 255 Shimo-Okubo, Sakura, Saitama 338-8570}}
\affiliation{\textsuperscript{16}\textit{Department of Physics, Tor Vergata University of Rome, Via della Ricerca Scientifica 1, I-00133 Rome, Italy}}
\affiliation{\textsuperscript{17}\textit{INAF - Astronomical Observatory of Rome, Via Frascati 33, I-00078 Monte Porzio Catone (Rome), Italy}}
\affiliation{\textsuperscript{18}\textit{Department of Astronomy, University of Maryland, College Park, MD 20742, USA}}
\affiliation{\textsuperscript{19}\textit{NASA/Goddard Space Flight Center, Greenbelt, MD 20771, USA}}
\affiliation{\textsuperscript{20}\textit{Department of Astronomy, Kyoto University, Kyoto 606-8502, Japan}
}

\title{Microcalorimeter X-ray Astronomy}

\Abstract{
Microcalorimeters have demonstrated success in delivering high spectral resolution, and have paved the path to revolutionary new science possibilities in the coming decade of X-ray astronomy. There are several research areas in compact object science that can only be addressed with energy resolution $\Delta E$\,$\ltsim$\,5\,eV at photon energies of a few keV, corresponding to velocity resolution of $\ltsim$ a few hundred km\,s$^{-1}$, to be ushered in by microcalorimeters. Here, we review some of these outstanding questions, focusing on how the research landscape is set to be transformed (i) at the interface between  accreting supermassive black holes and their host galaxies, (ii) in unravelling the structures of accretion environments, (iii) in resolving long-standing issues on the origins of energy and matter feedback, and (iv) to test mass-scaled unification of accretion and feedback. The need to learn lessons from \hitomi\ and to make improvements in laboratory atomic data precision as well as plasma modeling are highlighted.
\vspace*{0.2cm}
}

\begin{document}
\fontfamily{lmss}\selectfont
\flushbottom 

\maketitle 

\section{Introduction}
\vspace*{0.1cm}

Cosmic X-ray astronomy turns 60 this year \cite{giacconi62}. In this relatively short period, X-ray studies have transformed our view of the hot and energetic side of the universe. But the field still lacks capabilities that are commonplace in other domains. While astronomers have been splitting optical photons into their respective energies for more than a century using prisms and gratings with effective resolving powers ($E/\Delta E$) of several thousand and more, similar capabilities still do not exist on X-ray telescopes in orbit.

The above threshold in spectral resolution is important on physical grounds. Much of the gas in the cosmos, in planetary systems, stars, the interstellar medium and some clusters of galaxies, shows characteristic motions of order $\sim$\,100s\,km\,s$^{-1}$ \cite{branduardiraymont13, hitomi_perseus}. These motions can only be resolved and mapped if $E/\Delta E$\,$>$\,1,000. A number of key plasma diagnostics including density, ionisation state, and metallicity, rely on isolating weak and often closely spaced emission lines \cite{hitomimetallicity}. This, again, requires adequate spectral finesse. Inflows and outflows around galactic nuclei are also often characterised by relatively mild speeds as they traverse the threshold of the central black hole's gravitational sphere of influence  \cite{fabian12araa}. 

Detecting and splitting X-ray photons is non-trivial. A typical $\sim$\,10\,keV X-ray photon has a wavelength about a thousand times smaller than an optical one. Detecting these requires exceptionally precise optical surface smoothness, and small incidence-angle focusing optics \cite{sewardcharles}.
Absorption and scattering in the Earth's atmosphere necessitates X-ray telescopes being space-based. X-ray gratings can deliver the requisite spectral resolution needed, but currently only at low energies \cite{letgs, hetgs, denherder01}. Most such X-ray observations are photon-starved, and also unsuited to observations of extended objects in which spatial and velocity information gets mangled.

But cosmic X-ray spectroscopy is about to enter a new era through the use of microcalorimeters. These devices can measure the heat deposited in an absorption layer by individual X-ray photons. When cooled to cryogenic temperatures of $\sim$\,50\,mK, the detectors achieve low thermal noise and high energy sensitivity \cite{porter10, mitsuda10}. These demanding requirements make them challenging to operate in space, but we can expect $\frac{E}{\Delta E}$\,$\gtsim$\,1,000 for astrophysically important spectral signatures spanning the energy range of $E$\,$\sim$\,5--10\,keV with the upcoming generation of X-ray missions \cite{xarmmission, athena}. Being non-dispersive, microcalorimeters optimise high energy resolution together with high quantum efficiency for both point-like and extended cosmic sources. 

The \hitomi\ mission already gloriously demonstrated the potential of microcalorimeter science in-orbit, uncovering a surprisingly quiescent intergalactic medium in the Perseus cluster of galaxies through measurements of hot plasma spectral features with an exquisite precision of 10\,km\,s$^{-1}$ \cite{hitomi_perseus}. But the mission had barely begun when it suffered a premature demise \cite{hitomimission}. This followed prior ill-fated attempts to place other calorimeters in orbit including {\em Astro-E} and XRS/{\em Suzaku} \cite{xrs,gandhi18comment}. Paving the road to high spectral resolution X-ray astronomy has proved to be exceptionally arduous. At energies above a few keV, in particular, current grating spectrometers lose sensitivity, and X-ray astronomy has never seen a mission exploit this energy regime at high resolution, other than \hitomi. This is set to change with \xrism's launch around 2023  \cite{xarmmission} and 
\athena\ \cite{athena} in the early 2030s, optimised for energies $E\sim$\,0.5--10\,keV and spectral resolution $\frac{E}{\Delta E}$\,$\gtsim$\,1,000 at some of the key line transition energies. A few key design parameters for these missions and their respective spectrometers are listed in Table\,\ref{tab:missions}, and compared against the current best spectral resolution available with gratings on \chandra. 

\begin{table}[]
    \caption{\textcolor{teal}{\textbf{Salient nominal parameters of current and approved high spectral resolution X-ray instruments sensitive at 6--7\,keV.}}}
    \begin{tabular}{c|c|c|c}
    \hline
    \hline
    Instrument  & Ref. & $R$\,=\,$\frac{E}{\Delta E}$ & Area (cm$^2$)\\
    \hline
    \hline
    \chandra{} HETGS  & \cite{hetgs} & 165 & 28\\
    \xrism{} Resolve  & \cite{xarmmission} & 1,200 & 210\\
    \athena{}$^\dag$ X-IFU   & \cite{barret18} & 2,600 & 1,300\\
    \hline
    \end{tabular}
    ~\\
    \textcolor{mypurple}{The Spectral Resolution $R$ and the effective area listed are for the Fe\,K ($\approx$\,6--7\,keV) energy range. The first-order grating parameters are listed for HETGS. Values of $\Delta$$E$\,=\,5\,eV and 2\,eV are assumed for Resolve and X-IFU, respectively, as nominal goals. $^\dag$At the time of writing, a  redefinition exercise is ongoing, to consider optimised science possibilities with a streamlined mission designated {\textit{NewAthena}}.}
    \label{tab:missions}
\end{table}

Here, we review the game-changing advances expected with microcalorimeters for accreting compact objects. Accretion of matter is the primary mechanism thought to drive the steady growth of black holes in stellar-mass X-ray binary systems (XRBs) as well as in supermassive black holes in active galactic nuclei (AGN), as opposed to the more convulsive merger-driven growth that we are  beginning to probe with gravitational-wave discoveries \cite{ligo}. 

\begin{figure*}[htp]
\centering
\includegraphics[width=16cm]{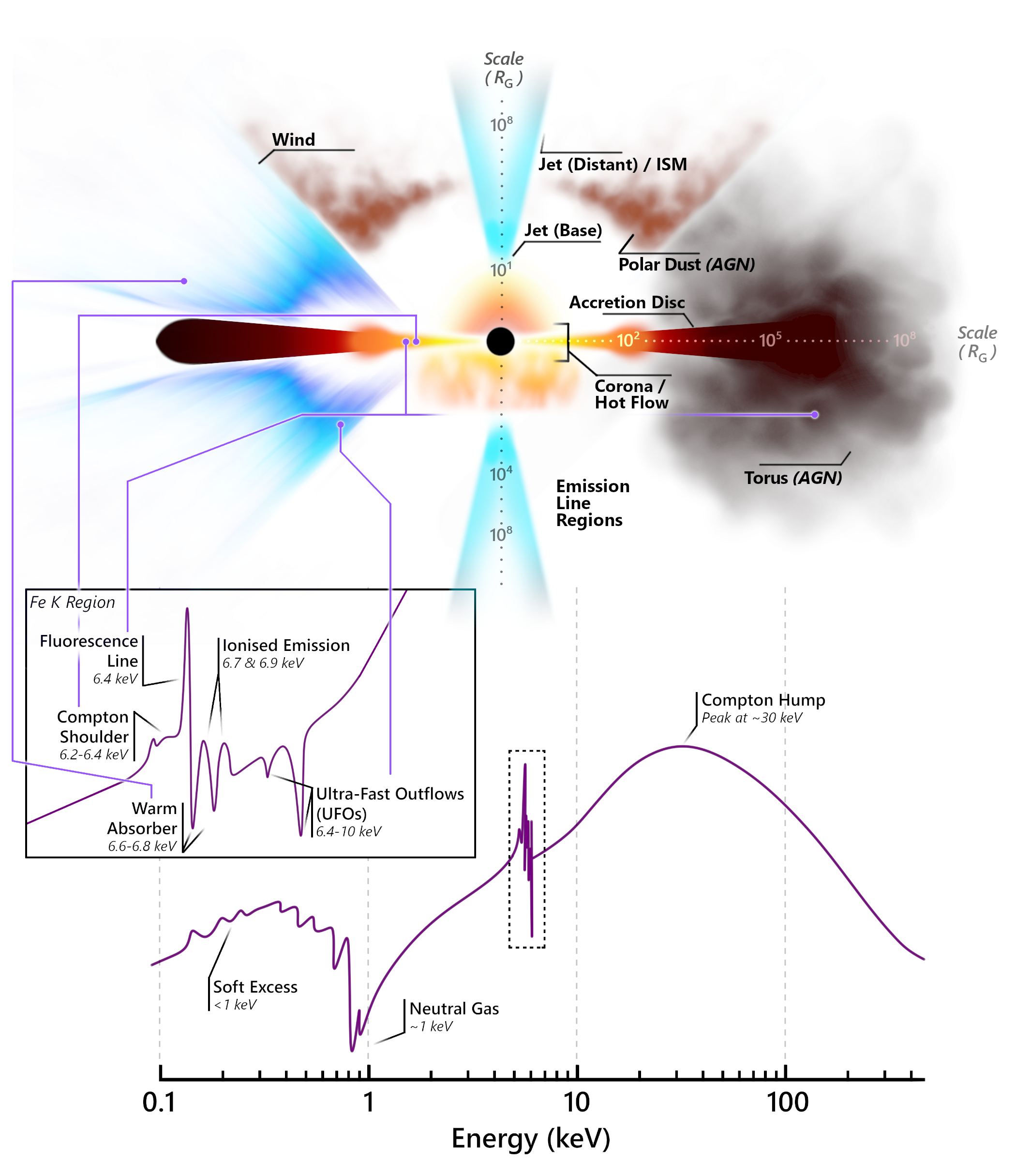}
\caption{\textcolor{teal}{Schematic overview of the environments of accreting compact objects and their X-ray spectra.}  \textcolor{mypurple}{Many components in the inner accretion regimes are common to both AGN as well as XRBs, including the accretion disc, corona/hot flow, and a relativistic jet. The left vs. the right-hand side depict differing potential models of the outskirts that may be attributed to differing physical conditions (e.g. some components exclusive to AGN are labelled on the right-hand-side). The geometry of several components remains unclear (e.g. the corona is depicted with a spherical vs. an inhomogeneous pancake-like geometry above and below the nucleus, respectively, for illustration) and the origin of others controversial (e.g. `the soft excess'). The `Emission Line Regions' indicate the broad and narrow line regions seen in AGN and nebular regions observed in XRBs. Approximate scales are denoted in Gravitational Radii ($R_{\rm G}$\,=\,$\frac{GM}{c^2}$) extending along the principal axes. The physical components manifest in various spectral signatures depicted at the bottom, and discussed throughout the text. The inset zooms-in on the Iron K energy range around 6--7\,keV.}
}\label{fig:overview} 
\end{figure*}

There are many unifying aspects to the structure of the accretion zones across the full spectrum of compact object masses. Fig.\,\ref{fig:overview} attempts to illustrate this unification in terms of the physical structures and spectral signatures seen across the stellar- and super-massive compact object regimes. In order for a particle to accrete onto a central compact object, it must traverse a huge dynamic range spanning several orders of magnitude in physical scales (and corresponding velocities). If accretion really is scale-invariant, it should be possible to illustrate all compact object environments in terms of approximate mass-scaled size units, as in Fig.\,\ref{fig:overview} where scales are denoted in Gravitational radii ($R_{\rm G}$\,=\,$\frac{GM}{c^2}$, with $G$ the gravitational constant and $c$ the speed of light). 

\begin{figure*}[h]
\begin{minipage}[hb]{18cm}
\begin{minipage}[b]{11.5cm}
\hspace*{-0.3cm}\includegraphics[width=11.75cm]{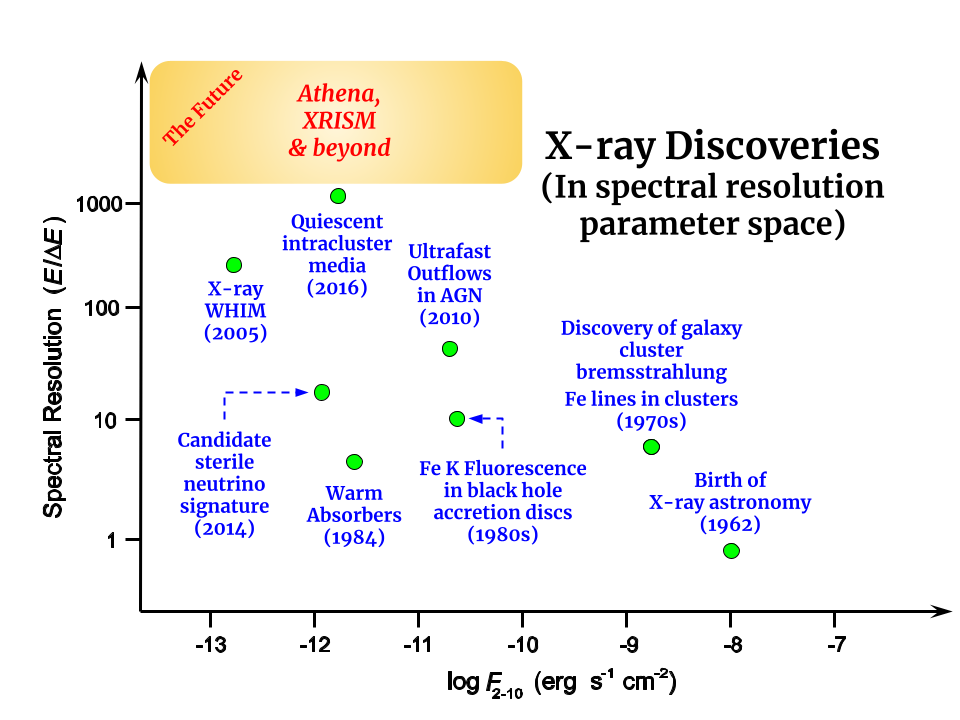}
\end{minipage}
\begin{minipage}[b]{6cm}
\caption{\textcolor{teal}{The road to key discoveries in X-ray astronomy as a function of increasing spectral resolution and improving continuum sensitivity, inspired by Harwit \cite{harwit81}}. \textcolor{mypurple}{The relevant key discoveries from right to left are from Giacconi et al. \cite{giacconi62}, Gursky et al. together with Mitchell et al. \cite{gursky71, mitchell76}, Barr et al. \cite{barr85}, 
Tombesi et al. \cite{tombesi10}, Halpern \cite{halpern84}, Hitomi Collaboration \cite{hitomi_perseus}, Bulbul et al. \cite{bulbul14} and, finally, Nicastro et al. \cite{nicastro05}. Source fluxes in the 2--10 keV band (or equivalent  continuum sensitivities in 100\,ksec of exposure) are denoted on the x-axis, and spectral resolution on the y-axis. Unexplored parameter space (at energies $E$\,$\gtsim$\,6\,keV relevant herein) is indicated by the shaded golden area.  \newline}
\label{fig:parameterspace}}
\end{minipage}
\end{minipage}
\end{figure*}

But not all aspects of such unification attempts are without controversy, as we will discuss herein. Accretion physics at all scales will thus strongly benefit from gains in spectral resolution, in order to resolve these controversies and break new ground in accretion studies. No clean, high quality microcalorimeter spectrum of an accreting source has yet been obtained, so this field is rife with discovery potential, as history has repeatedly demonstrated. Fig.\,\ref{fig:parameterspace} highlights this point by indicating key astronomical discoveries that have been enabled by increasing X-ray spectral resolution and flux sensitivity. 

We focus herein on four key questions examining the structures and physical conditions of accreting black hole environments (Section\,2), and the nature of energy and matter feedback (Section\,3). This should not be read as an exhaustive review, rather a taster of the low-hanging fruit waiting to be plucked in high spectral resolution studies. Space constraints preclude detailed discussions of themes such as the nature of Ultraluminous X-ray sources, tidal disruption events, accreting neutron stars, and investigations of exotic transients.

\section{The Accretion Environment}

\vspace*{0.5cm}
Microcalorimeter science will dramatically widen our reach of the {\em physical scales} that can be probed around accreting compact objects, as well as the dynamical range of their {\em physical conditions} amenable to study. We begin by focusing on two key questions encompassing both these aspects:\\
\vspace*{-0.5cm}
\begin{enumerate}[label={\textbf{[\arabic*]}}]
    \item {\textbf{How does matter accrete from large scales?}}
     \vspace*{-0.7cm}
    \item {\textbf{What are the physical conditions in the inner vicinities around compact objects?}}
\end{enumerate}
\vspace*{0.5cm}

\noindent 
\textbf{2.1.  Gas flows at the interface between the nucleus and host galaxy:} The Keplerian velocity of a particle of interstellar matter located at the radius of influence $r$ of a massive black hole is $v$\,$\sim$\,100$\sqrt{(M/10^7\,{\rm M}_{\odot})(4\,{\rm pc}/r)}$\,km\,s$^{-1}$. Resolving such `mild' velocities requires $E/\Delta E$\,$>$\,few 1,000. The inability to probe these velocities in X-rays has hampered our understanding of the physical state and kinematics of hot gas at the interface between the black hole and its host galaxy. 

Gas needs to shed several orders of magnitude of angular momentum if it is to be accreted from here onto the central engine. The physical mechanisms driving these \underline{gas inflows} are thought to be gravitational instabilities, triggered either secularly or externally. Secular processes can include bending modes and warps arising in lopsided gas and nuclear stellar discs \cite{hopkins12}. On larger scales, stellar bars might bring in fresh supplies of gas from the host galaxy \cite{cisternas15}. External processes comprise galactic mergers and interactions, whose role still remains controversial and may be dependent on luminosity and redshift \cite{marian19, storchibergmann19}. Microcalorimeters will enable measurements of faint extended gas flows with typical plasma speeds of $\sim$100s km\,s$^{-1}$ within the hot bulges and haloes of the nearest galaxies.

Gas inflows not only feed, but also \underline{obscure} the central engine from direct view. The demographics of `obscured' AGN require that nuclear veiling gas must be distributed in a geometrically thick, axisymmetric configuration, generically referred to as the `torus' (Fig.\,\ref{fig:overview}). 
Its presence is key to unifying the characteristics of all AGN under a simple inclination-dependent metric, at least to first-order \cite{antonucci93, netzer15}. In X-rays, the torus manifests as line-of-sight absorption as well as line and continuum scattering signatures, and can explain the integrated shape of the cosmic X-ray background radiation \cite{gilli07, ueda14, ananna19}. Yet, its structure and dynamics remain much debated: e.g. is the gas in the toroidal zone undergoing bulk Keplerian rotation, or is it instead dominated by outflows? How is this geometrically thick configuration supported and what is the cloud filling factor \cite{elvis00,  markowitz14, hopkins16, ramosalmeida17}? 

The canonical torus picture envisions strong reprocessed thermal infrared reemission from obscuring clouds in the {\em equatorial} plane. Recent interferometric observations have instead revealed a surprisingly different spatial distribution, with dust being found to be elongated along the {\em polar} direction on scales of a few parcsec (Fig.\,1), likely driven outwards by radiation pressure \cite{Hon13}. The dynamical state of cold gas coupled to this dust has been studied with ALMA \cite{Izu18,gatos1}, finding complex flows with characteristic velocities of a few 100\,km\,s$^{-1}$. The  characteristic inmost scale size of the torus as inferred in the infrared is set by radiative physics of dust sublimation \cite{barvainis87}, but X-ray absorbing gas must span this boundary and exist within the sublimation radius down to scales of the accretion disc. Microcalorimeters will be capable of probing \underline{hot} \underline{gas kinematics spanning the dust sublimation radius}.

X-ray microcalorimeters will allow simultaneous studies of the three fundamental phases of interstellar matter -- \underline{dust, atomic gas, and ionised gas} -- in particular, high $Z$-number gaseous elements that ALMA cannot probe. High resolution X-ray spectroscopy offers the capability to decouple these phases, and to track all ionisation states of many high-$Z$ elements, something not possible in other wavebands. With sufficient signal, dust absorption and scattering features could be isolated from gas, delivering estimates of grain size, composition and elemental depletion around the dense environments of accreting sources \cite{rogantini18, corrales16}. 

The most prominent line transitions in the X-ray regime are associated with Iron (Fe). A relatively high cosmic abundance makes the \underline{Fe emission and absorption features} excellent spectral probes of gas physical conditions, geometry and dynamics \cite{basko78, fabian00_fe}, including the torus \cite{molendi03, mytorus, balokovic18, shu10}. Any coherent Keplerian flows should be easily resolved as a result of rotational broadening of the Fe K$\alpha$ and K$\beta$ fluorescence line doublets around 6.4\,keV and 7.1 keV, respectively, and the inner edge of the gas torus localised to better than 3\% uncertainty in nearby systems (assuming that Keplerian flows at the inner edge dominate the reflected emission \cite{Rey14}). 

    The only microcalorimeter AGN spectrum obtained so far is that of NGC\,1275, the low luminosity nucleus of the Perseus cluster \cite{hitomi_3c84}, where a narrow Fe line core (with characteristic velocity $v$\,=\,500--1,600\,km\,s$^{-1}$) was found by \hitomi, tracing reflection in a putative extended torus out to scales of a few hundred pc. This single observation is already in \underline{conflict with prior (grating) studies} of more luminous Seyferts and quasars with \chandra, where an Fe\,K$\alpha$ line width of $v$\,$\approx$\,2,000--3,000\,km\,s$^{-1}$, and much broader in some cases, suggested an origin within the torus dust sublimation radius \cite{shu10, minezaki15, Gan15, uem21}. This could be new evidence of luminosity (or accretion-rate) dependent evolution of the obscuring geometry \cite{ezhikode17, ricci17}, since NGC\,1275 has historically been a weak AGN but with a recent variable, brightening trend \cite{fabian15_ngc1275}. Spatial broadening may also have biased prior grating studies of Seyferts as a result of extended Fe line emission, but the magnitude of this bias remains unclear \cite{Liu16, uem21}; spatial extents equivalent to tens to hundreds of pc, orders of magnitude larger than the inner torus, would be required \cite{masterson22}.
    
    Microcalorimeter (non-dispersive) spectra of more \lq standard\rq\ AGN, free of contamination by cluster emission, are necessary if we are to scrutinise this putative luminosity dependence and measure gas motions on inner torus scales. Such observations ought to be high priority for \xrism\ and \athena\ upon launch, and can test the hot X-ray gas distribution against the distribution of cooler gas and dust components inferred at other wavelengths in local samples of AGN \cite{gatos2, Esparza-Arredondo2021}. Sensitive studies of Fe\,K$\alpha$ emission on {\em extended} $\sim$\,kpc scales will also become possible in nearby AGN hosts with \athena{}, tracing the reflection echoes of long-term nuclear activity and improving upon photon-starved studies with {\em Chandra} \cite{marinucci12, andonie22}.

Going beyond keplerian flows, high-order grating data from \chandra\ show apparent blueshifts in fluorescent Si emission lines at 1.74 keV of a few hundred km\,s$^{-1}$, which may be evidence for \underline{outflowing gas} in a polar elongated region, an important test of the \lq polar dust scenario\rq\  \cite{Liu19, shu10}. But this evidence is still indicative at best, and such studies have been limited to a handful of the brightest, nearby AGN. Sub-mm observations also predominantly find equatorial, as opposed to polar, outflows \cite{gatos1}. X-ray line features from the inner zones of any polar outflow are needed to assess this model critically. 
    The Resolve microcalorimeter on \xrism\ will have a spectral resolution $\frac{E}{\Delta E}$\,$\approx$\,1,200 at the prominent fluorescence Fe line energy of 6.4\,keV, and its low background is expected to robustly uncover outflows down to levels several times fainter than currently possible (Table\,\ref{tab:missions}; \cite{guainazzitashiro18, xrismscienceteam_wp_2020}). \athena's desired combination of high spatial resolution of $\approx$\,5--10\,arcsec together with an $\frac{E}{\Delta E}$\,$\gtsim$\,2,600 for the X-IFU instrument \cite{barret18} will greatly expand this science by tracing motions of X-ray reflecting clouds in nearby galactic bulges, and studying the complex interplay between star-formation and AGN activity \cite{Kaw19, gatos1, athena}.

The optical depth and ionisation state of obscuring clouds will also be laid bare through the finesse of microcalorimeters. A \underline{Compton shoulder (CS)} to the neutral Fe line is expected between 6.2--6.4 keV as a result of Compton scattering in cold matter, with a strength determined by gas optical depth and covering factor \cite{matt02}. \xrism\ is expected to find this feature in many AGN, improving upon a few tentative detections thus far \cite{iwasawa97, bianchi02}, and so provide precise constraints on cloud geometry and optical depth when combined with broadband modelling. \underline{Chemical composition} constraints will be possible through accurate metallicity determinations of the Fe edge, and by measuring the the Nickel and Iron fluorescence line strengths relative to each other and to the CS \cite{molendi03}. This will shed light on the origin of the accreting material, e.g. the starburst history in the case of AGN tori \cite{hikitani18}, or abundances of material that is accreting from the donor star in XRBs \cite{watanabe03}. Fe abundance has been found to be super-Solar in the bright Circinus Galaxy  ($A_{\rm Fe}$\,=\,1.7) with $\approx$\,10\,\% precision \cite{hikitani18}. Microcalorimeters will improve upon this, and will cast the net of detection possibilities much wider across the AGN population. 

We have discussed probes of kinematic and of gas physical conditions above. Some of the most transformational science is likely to come from the ability to provide \underline{{\em simultaneous} constraints} on these, as we illustrate with the following example. The ionization state of the absorbing and reflecting gas at any distance $r$ from a central engine is quantified by an `ionization parameter' ($\xi$\,=\,$L$/$nr^2$) dependent on the ionizing power $L$ and the local space density $n$. For every ionic stage of Iron, the transition energy centroid $E$ ramps up by a few tens of eV (up to Fe\,{\textsc{xvii}}, beyond which $E$ increases even faster \cite{Kall82}). This corresponds to log\,$\xi$ ramping up approximately by 0.1--0.2 dex at each step \cite{Nag89}. Thus, $\xi$\ can be measured in warm plasmas to within a precision of 0.1\,dex, if photon energies around the Fe band of $E$\,$\sim$\,6--7\,keV can be  constrained to an accuracy $\lesssim$\,5--10\,eV (i.e., a velocity resolution $\Delta v$\,$\lesssim$\,200--500\,km\,s$^{-1}$).  

We tested this capability of jointly constraining $v$ and $\xi$ by simulating the presence of a putative (and so-far undetected) ionised outflow in the prototypical Seyfert galaxy NGC\,4388. This is amongst the brightest of nearby, hard X-ray Compton-thin (mildly obscured) AGN with a prominent, narrow fluorescence line \cite{bat105}. Prior studies of this galaxy are used to inform our base X-ray model comprising a power-law and neutral emission line components \cite{Shi08}, atop which we introduced an additional putative ionised outflow, with a modest blueshifted velocity $v$\,=\,--300 km\,s$^{-1}$. The spectrum of the outflow itself is modelled as an ionised reflection component using the {\textsc{xillver}} code \cite{Dau14,Gar14} with $\log \xi$\,=\,1.8, corresponding to Solar metallicity gas with a volume density of $n_{\rm H}$\,=\,10$^{6}$ cm$^{-3}$ located at a distance of 0.1\,pc from a central engine with ionizing luminosity $L$\,$\approx$\,$10^{43}$\,erg\,s$^{-1}$. This could represent the optically-thick base of the dusty polar outflow known to exist in this source \cite{asmus16}. Its simulated strength has been constrained to be consistent with flux limits inferred from prior X-ray  observations \cite{Shi08}. 

\begin{figure}
\centering
\hspace{-.30cm}
\includegraphics[angle=-90,width=4.52cm]{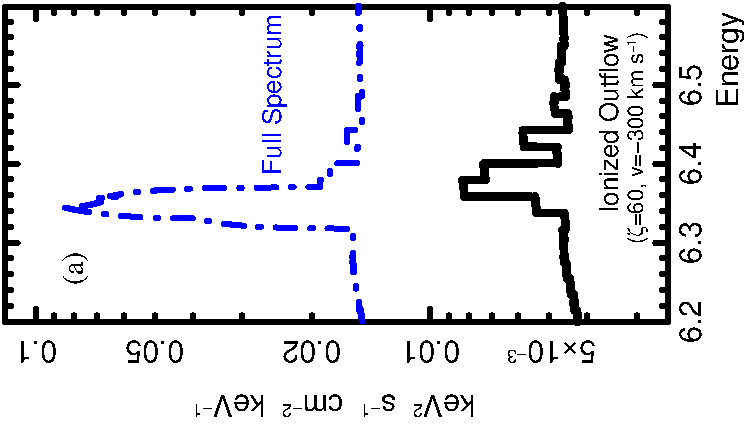}
\hspace*{-0.3cm}
\vspace*{0.cm}
\includegraphics[angle=-90,width=4.2cm]{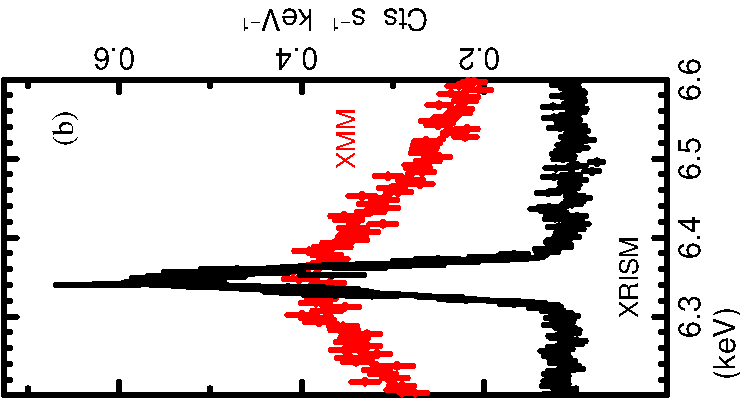}
\vspace*{0.5cm}
\hspace{-0.5cm}
\includegraphics[width=9cm]{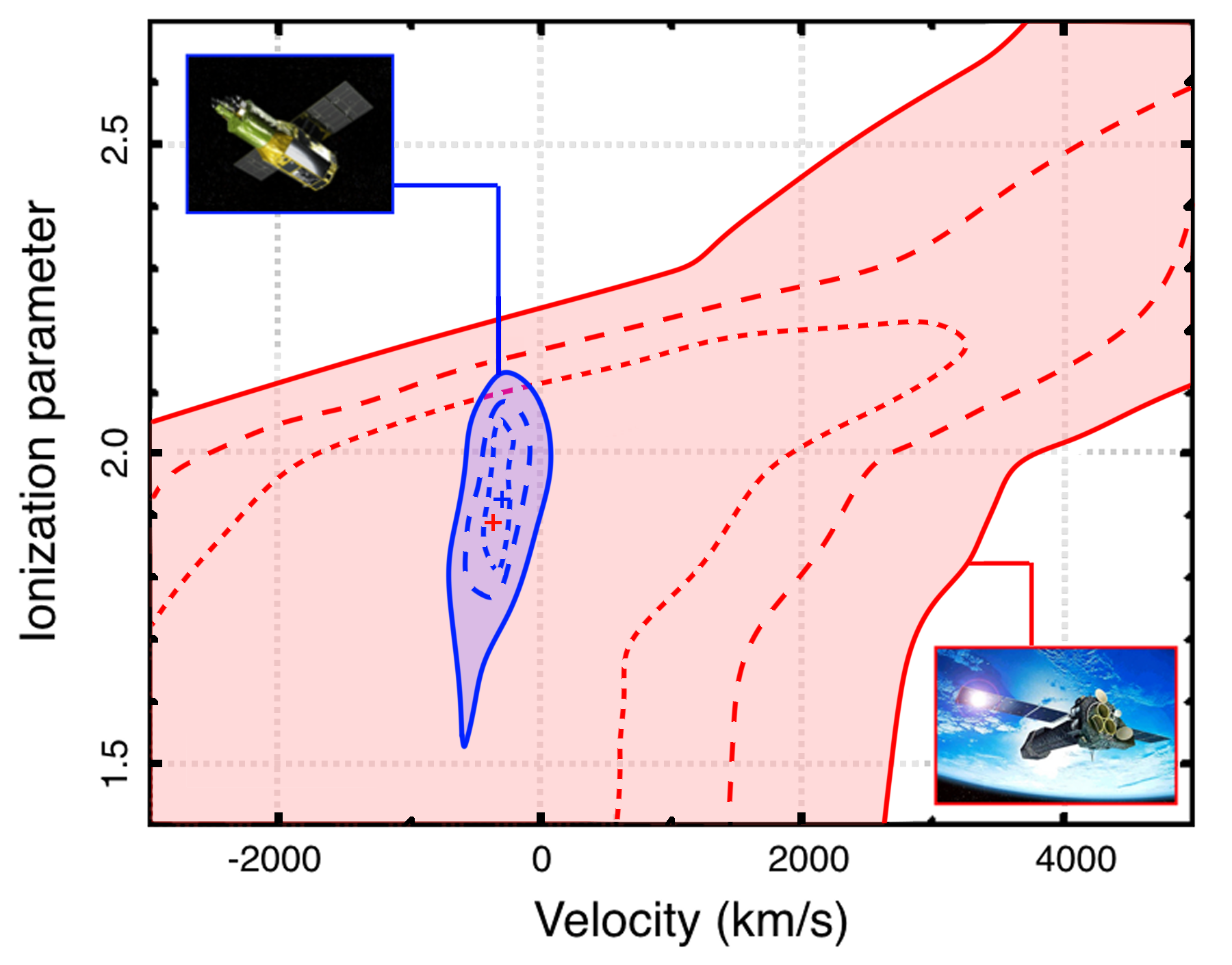}
\caption{
\textcolor{teal}{{\em Comparison of joint physical constraints on a mild ionised outflow with a microcalorimeter vs. a CCD.}} \textcolor{mypurple}{\underline{(a):}
X-ray spectral model of the nearby Seyfert NGC\,4388, including an additional ionised outflow in the Fe band. The ionised outflow contributes only a small fraction of the full spectrum line flux. \underline{(b):} 300\,ksec  \textit{XMM-Newton} (CCD; red) and \xrism\ (microcalorimeter; black) simulated counts spectra. A zoom-in around the Fe lines is shown, while the fit itself has been computed over a wider energy range.  \underline{Lower panel:} Fit contours at 68\%, 90\%, and 99\% confidence levels between the ionization parameter log\,$\xi$ and the outflow velocity. The red and blue contours are for \textit{XMM-Newton} and \xrism, respectively. Negative velocities indicate blueshifts. The Resolve microcalorimeter on \xrism\ will be  sensitive enough to decouple the degeneracy apparent with the CCD, and to pick out fine velocity shifts that are physically relevant in the outer ramparts of the accretion regime. 
}\label{fig:cont}}
\end{figure}

Spectra simulated with 300\,ksec--long CCD (\xmm\ EPIC) and microcalorimeter (\xrism\ Resolve) observations are shown in Fig.\,\ref{fig:cont}. The simulated CCD emission line is considerably broader than the corresponding line as seen by the microcalorimeter, and is even broader than the plotted energy range. The bottom panel shows the best-fit confidence contours between log\,$\xi$ and $v$, derived by fitting the 6.2--6.6\,keV spectral window. Degenerate coupling between the two parameters limits the constraints possible with the CCD, but not for \xrism, demonstrating the power of the microcalorimeter to robustly isolate blueshifts from ionisation effects. The structured Fe line complex together with the high spectral resolution of Resolve provide this distinguishing capability. $\xi$ will be constrained to a precision of better than 10\%. The uncertainty on $v$ is $\approx$\,70\,km\,s$^{-1}$, more than a factor of 20 better than what is possible with CCDs, even in this conservative simulation. Though we have focused on the Fe\,K$\alpha$ regime here, the K$\alpha$ to K$\beta$ line strengths ratio can also be leveraged to provide further constraints on $\xi$ \cite{Rey14}. \athena\ should improve upon \xrism's best spectral resolution by a further factor of $\approx$\,2--4 in the Fe band with its integral field instrument X-IFU \cite{barret18}, allowing additional physical conditions such as abundance studies of the material outflowing from the torus. 

As one final line of investigation with no current observational constraints, we mention the unknown prevalence of \underline{binary supermassive black holes} (SMBHs). Such objects in the process of merging are expected to be prime targets for the next generation of space gravitational wave interferometers \cite{hughes02}. If one or both of the component SMBHs are accreting, complex reflection line profiles are expected \cite{Yu01, jovanovic20}. The combination of electromagnetic line shifts with gravitational wave detections will enable unique constraints on relativitic physics and SMBH growth models via accretionBinary SMBHs that have suffered a recoil of magnitude $\sim$\,500\,km\,s$^{-1}$ or more \cite{komossa08b} could be easily isolated, if the recoiling SMBH is accreting and shows X-ray line features clearly offset from  narrow lines that arise farther out. There is also the potential to learn how nature circumvents the so-called `final parcsec problem' -- the unknown mechanism by which binary black holes can lose angular momentum and thereby overcome the final hurdle to merging \cite{armitagenatarajan02}.

\vspace*{1cm}
\noindent
\textbf{2.2.  Spectrally resolving the inner accretion regime:} The hottest plasmas at the hearts of accreting sources are key to testing the behaviour of matter under extreme gravity. Their studies open the gateway to measuring several fundamental physical quantities including (1) central object mass $M$, (2) its spin magnitude $a$ and direction, together with (3) the dynamics and (4) the geometry of the inner accretion regimes. There appear to be some beautifully simple homologies between the stellar-mass and supermassive scales (as illustrated in Fig.\,\ref{fig:overview}), though several of these correspondences remain controversial \cite{done07}. Studies of both classes of sources are thus enlightening and important.  

The inner accretion zones are known to undergo dramatic flux and spectral transitions during well-defined accretion episodes, or `outbursts,' in XRBs. Several distinct transient \underline{accretion states} have been identified in these outbursts, broadly classified as `hard', `soft' and `intermediate' based upon the X-ray continuum spectral slope, and flux variability amplitude  \cite{remillardmcclintock06, done07}. These are thought to be triggered by instabilities in the accretion flow driven, ultimately, by accretion rate variations in the outer disc \cite{lasota01}, though the detailed processes underlying the state transitions, and associated changes in geometry and physical state, remain to be clarified. Longer transition timescales in AGN preclude a proper understanding of accretion states in supermassive black holes. Recent years have seen an increasing number of identifications of so-called `changing-look' or `changing-state' AGN that switch between AGN classes \cite{shappee14}. However, their switchover timescales are far too rapid to be explained solely by viscous accretion rate changes. Irradiation effects, magnetic support, and external perturbations such as tidal disruption events may play a significant role in explaining their rapid transitions \cite{macleod16, dexterbegelman19, ricci20}. 

Dynamical motions deep in the gravitational potential wells of compact objects can reach significant relativistic speeds. Low spectral resolution $\frac{E}{\Delta E}$\,$\ltsim$\,100 thus easily suffices for the most rapid motions, as pioneering studies demonstrated more than 20 years ago \cite{fabian89, tanaka95}. Particularly in a disc-dominated accretion state, the \underline{inner edge of the disc} likely lies close to or reaches the innermost stable circular orbit (ISCO), a key dynamical threshold in the space-time metric of black holes \cite{sewardcharles}. Inner disc reflection of primary coronal X-ray emission manifests as Fe fluorescence emission  broadened due to Doppler and relativistic effects \cite{georgefabian91}. The magnitude of broadening depends on the relative geometrical configuration and extent of the disc versus the corona and, crucially, scales with $a$ \cite{laor91, brenneman13}. Because of their ability to constrain the innermost disc zones, broad Fe line studies have matured over the past thirty years and now occupy an important niche within X-ray astronomy \cite{fabian00_fe, dovciak04, Gar14, rey21}.

Progress has been impressive enough that the focus has shifted from reducing statistical uncertainties to deliberations on \underline{systematic modelling issues}. For instance, debates rage regarding the proximity of the inner disc edge to the ISCO in other accretion states \cite{miller06_swiftj13753, donediaztrigo10, turnermiller09, hagino16}, and how Fe line constraints match up with those inferred from reverberation mapping and disc continuum modelling techniques \cite{mcclintock06, yamada09, uttley14, demarco21_maxij1820}. It has also been postulated that it is not the inner disc, but rather the characteristic scale height of the corona, that instead varies with accretion rate \cite{kara19}. The line strength depends upon the disc inclination angle, and there remain disagreements between the inferred inner disc inclination (as measured in X-rays) vs. the binary orbital plane (as inferred in the optical) \cite{connors19, draghis21}. This may point to warped inner disc geometries, which would then need to be accounted for self-consistently when inferring $a$ \cite{abarr21}. The origin of the so-called `soft excess' below 1\,keV (Fig.\,\ref{fig:overview}) also remains under discussion, with blurred reflection, thermal Comptonisation, and models including contamination due to ionised host galaxy gas having been proferred \cite{crummy06,noda13,petrucci18}. Resolving these disagreements has important consequences for cosmic spin measurements, black hole growth, and comparisons between XRBs and for black hole binary mergers events detected by LIGO/Virgo \cite{rey21, Belczynski21}. The relevance of spin in powering relativistic jets also remains controversial \cite{narayan12, russell13}, partly a consequence of small numbers of accurate spin measurements thus far. 

Many of these issues can be traced back to the difficulty of properly decoupling broad reflection and absorption features from the \underline{underlying X-ray continuum}. The continuum itself may also be more complex than a simplistic single thermal plasma, with hybrid plasmas and multi-temperature zones plausibly being more physically realistic \cite{coppi99, makishima08,yamada13}. Since most of the radiated accretion power emerges in the X-ray continuum, at least in XRBs, accurately characterising its origin is critical. 

If we are to break systematic degeneracies, boosting detected counts alone will not suffice. Higher spectral resolution is critical for \underline{proper deconvolution} of confusing narrow features, absorption lines, reflection edges and other subtle spectral curvature. As an additional example, Fabian et al. \cite{fabian20_abs} demonstrate the possibility that {\textsl{absorption}} of reflected photons within a disc atmosphere around 8--9\,keV can mimic the appearance of a broad blueshifted outflow in the well-studied narrow-line Seyfert 1 galaxy IRAS\,13224--3809. If true, this will allow studying not only the disc atmosphere, but also disc dynamics, within a few $R_{\rm G}$ of the black hole through studies of absorption features. Microcalorimeter spectra could resolve the shape of the absorption feature from the underlying continuum, locating and mapping the disc absorber precisely.

More generally, there is an enormous disparity in the dynamic range of physical scales over which emission and absorption operate. In the deep potential around a rapidly spinning black hole with $a$\,$\xrightarrow[]{}$\,1, most of the observed energy release from a luminous disc occurs very close to the centre ($\sim$\,50\,\% of energy is liberated within $<$5 $R_{\rm G}$; \cite{thorne74}). If this also applies to the coronal emission then it will be similarly compact. By contrast, warm absorption, outflows and dusty gas might arise on disc, torus, or galactic scales, with \underline{absorption along many sight-lines} that can impact the appearance of that inner emission. Microcalorimeters can make an important contribution in disentangling these. In particular, samples of \lq bare\rq\ AGN, with minimal absorption \cite{walton13}, need to be compared with more complex ones. \xrism\ can place the best constraints on just how bare such AGN are, and how well we can isolate the inner emission to robustly measure spin  (e.g. \cite{brenneman13}). \athena's high sensitivity will push this envelope to include large ensembles of targets at high redshift. $k$-corrections will redshift key emission and absorption features to lower observed-frame energies where X-IFU's effective area will peak \cite{barret18}. For example, mean uncertainties on spin $a$ under conservative assumptions could be $<$\,0.05 \cite{barret19}. Ideally, this would be further combined with broadband spectral sensitivity (e.g. with \nustar\ \cite{nustar} or future missions sensitive above $\sim$\,10\,keV), in order to accurately model the underlying continuum to high energies. This was highlighted in a recent study of one of the brightest X-ray transients MAXI\,J1820+070 \cite{kawamuro18}, where a $\sim$\,1\,keV blackbody \underline{emission-excess} has been reported that cannot be explained by disc continuum nor standard Fe fluorescence \cite{Fab20}. Instead, it is interpreted as thermal emission from a non-zero torque arising at the plunge region \textsl{inside} the ISCO where matter freely falls into the black hole \cite{Fab20}. Other transients may already have shown this feature \cite{Oda19}, but its faintness makes its detection and study non-trivial. Any additional ionised outflows superposed on this would render this near-impossible to isolate and interpret. Coordinated observations with microcalorimeters will thus be crucial to testing this scenario. 

Sources that happen to lie edge-on as seen from our vantage point will provide a unique line-of-sight to study the composition of their \underline{disc atmospheres}. In particular, for those seen at inclinations $\gtsim$\,80\,$^{\circ}$, scattering off extended plasma overlying the disc will dominate, without being blinded by radiation from the central accretion disc. Expected recombination X-ray emission should allow us to probe the plasma metallicity, density and ionisation structure \cite{jimenezgarate02}. Pushing down to energy resolutions $\Delta E$\,$\lesssim$\,10\,eV with microcalorimeters will open up the possibility to precisely measure radial stratification within the atmosphere, and also deconvolve double-peaked disc lines, something not possible with gratings. Only a few of these sources are currently known \cite{done14}, but \xrism\ and then \athena\ will push the flux limits to which these can be detected. 


Finally, it should be possible to detect motions of individual structures within the accretion stream (e.g. hot spots in the disc) in \lq real-time,\rq\ if (i) they are bright enough to stand out above the continuum, and if (ii) they display regular Doppler shifts. Tentative detections have previously been reported from Fe\,K--bright spots located at radial distances of just $\sim$\,10\,$R_{\rm G}$ in AGN;  \cite{iwasawa04}. Higher spectral resolution time-resolved observations will not only uncover secure examples, but also help to constrain the central compact object mass $M$ by modelling of the \underline{dynamical line modulation}.  

Dynamical line modulations could also be employed to obtain \underline{mass constraints} for XRBs. If the radial velocity modulation of the donor star, or of the accretor, can be measured as a function of orbital phase, this leads to an estimate of the binary mass function which, together with independent estimates of the inclination angle, constrains the masses of the individual XRB components. This proposed technique relies on detecting narrow absorption features in the disc wind of the accretor \cite{zhang12} or, complementarily, on emission line fluorescence from the surface of the donor star \cite{dashwoodbrown22}. The radial velocity sensitivity required is $\sim$\,10\,km\,s$^{-1}$, depending upon system parameters \cite{casares14}. This is within reach of microcalorimeters at the bright end, but it is still unclear whether such features can be routinely isolated and whether their point-of-origin within the binary can be constrained well enough to model the mass function. Prior detections of narrow features with \chandra\ are encouraging \cite{ponti18, torrejon10}, and if these techniques work, they have the potential to extend mass function measurements to the highly obscured population of XRBs in the Galaxy -- objects that have thus far eluded dynamical mass measurements in the optical \cite{casares14}.


\section{Energy and Matter Feedback}\label{sec:outflow}

\vspace*{0.5cm}
We next turn attention to the impact of accretion on the galactic-scale environment, a topic of intense interest in compact object physics, galaxy evolution and cosmology. Specifically, the following questions will shape our discussion below: 
\begin{enumerate}[label={\textbf{[\arabic*]}}]
    \item {\textbf{Are accretion and outflows scale-invariant?}}
     \vspace*{-0.3cm}
    \item {\textbf{What drives and regulates energy and momentum feedback?}}
\end{enumerate}
\vspace*{0.5cm}

\noindent
\textbf{3.1.  Scale-invariance:} Supermassive black holes at the centres of galaxies impact their host galaxies via radiation and outflows, the latter being in the form of collimated jets or wide-angle winds \cite{mcnamara07araa,fabian12araa,harrison17natas,morganti17frass}. Such energy and momentum `feedback' results in a symbiotic coupling of the central SMBH's growth to that of the host galaxy, as can be appreciated from the following simple energetic argument \cite{silkrees98, f99}: the gravitational binding energy of a typical galaxy is approximately $E_{\rm bind}$\,$\sim$\,10$^{60}$\,$(M_{\rm gal}/10^{12}\,{\rm M_\odot})^2/(R_{\rm gal}/{\rm 30\,kpc})$~erg. This turns out to be comparable to the rest-mass energy  associated with a growing SMBH, $E_{\rm rest}$\,$\sim$\,10$^{60}$\,($f_C/0.01$)($M$/10$^{8}{\rm M_\odot}$)~erg, if $f_C$\,=\,1\,\%, that is, 1\,\%\ of the rest mass energy can be fed back outwards, coupling to matter on galactic scales. Numerical simulations that have implemented simple energy coupling prescriptions between AGN output and circumnuclear gas have, indeed, succeeded in explaining some tight correlations observed between SMBH masses and host bulge properties \cite{ferrarese00, marconihunt03, dimatteo05}.

However, the processes transporting energy and metals from the nucleus out to interstellar and circumgalactic scales are complex, as is their net effect on the evolutionary history of the host galaxy. Therefore, it remains an open question whether jets or winds exert a larger influence on such evolution. An important limitation in determining the primary feedback channel is our lack of understanding of how jets and winds are launched, and high X-ray spectral resolution is key for advancing our knowledge of winds. 

Accretion disc winds are observed both in XRBs and AGN via highly ionised absorption and emission lines in moderate to high-resolution X-ray spectra \cite{diaztrigo16an,tombesi16an}. In AGN, winds span a wide range of velocities, from $\sim$\,100\,km\,s$^{-1}$ in the so-called \lq warm absorbers\rq\ observed mainly in the ultraviolet or soft X-rays \cite{halpern84, blustin05}, and reaching relativistic speeds in `Ultra-Fast Outflows (UFOs)' \cite{tombesi13mnras} of which the prototypical example is PDS\,456 \cite{reeves09apj}. UFOs are, therefore, expected to exert significant \underline{kinetic feedback}, and thus be important for host galaxy evolution, if confirmed to be at the origin of larger (parsec-scale) molecular outflows \cite{tombesi13mnras,feruglio15aa}. 

In contrast, most XRB warm absorbers do not show significant blueshifts and even winds in iconic black hole XRBs like GRO\,J1655--40 and GRS\,1915+105 show velocities $\lesssim$\,2000\,km\,s$^{-1}$ \cite{miller06nat,ueda09apj,neilsen12mnras,diaztrigo16an} (but see \cite{degenaar14apj,nowak19apj} for examples of larger velocities reported in some neutron star XRBs), and consequently, are not expected to contribute significantly to kinetic feedback. But when winds are present, the large {\em masses} expelled, of the order of 10$^{18}$~g~s$^{-1}$,  \cite{ueda04apj,ponti12mnras}, are comparable to mass accretion rates in these objects, and could result in depletion of the outer disc and a halt of accretion onto the compact object \cite{v404:munoz-darias16nature}, thus limiting the duty cycle of the XRB and ultimately determining the amount of global feedback.

Here, it is now important to reduce the uncertainties in the mass outflow rates, estimated as $\dot{M}_{\rm outflow} = \Omega\,n\,m_{\rm p}\,r^2\,v_{\rm out}$, where $m_{\rm p}$ is the proton mass and $n$, $r$, $v_{\rm out}$ and $\Omega$ are the outflow density, launching radius, velocity and solid angle, respectively. 
For example, the solid angle of the winds remains ill-constrained, but \underline{high detection rates} in AGN imply that winds must have wide-angled geometries, with solid angles covering at least 30\,\%\ of all sight-lines, and more after accounting for observational biases \cite{crenshaw03, tombesi13mnras}. In XRBs, there are important state dependences, with detections restricted (until recently) to equatorial sightlines in the soft state \cite{ponti12mnras}, though this paradigm may now be changing \cite{castrosegura21}. 

A major step in the unification of black hole accretion was the discovery of the fundamental plane of black hole activity relating radio and X-ray luminosities across more than six orders of magnitude in mass $M$ \cite{merloni03mnras,falcke04aa}, and indicating similar accretion and jet launching physics at all scales (cf. Fig.\,\ref{fig:overview}). Since winds, similar to jets, are a fundamental piece in the accretion--ejection paradigm, we might expect that they also scale with $M$. 

However, this \underline{scaling} remains largely unexplored (see \cite{king13apj,nomura17mnras,giustini19aa} for a few attempts), mostly due to the paucity of data and the difficulties of characterising the winds. The breakthrough that the field awaits is a better understanding of the primary drivers of wind launching. Microcalorimeters will tackle this in detail, as we discuss in the following section. Better spectral resolution and sensitivity will allow fair comparisons of stellar- and super-massive systems matched on physically meaningful parameters such as accretion state and the Eddington-scaled accretion rate. This will allow us to determine whether differences observed between the stellar and supermassive scales are due to different wind launching \textsl{physics} possibly varying with mass scale $M$, and to ascertain the impact of the wind physics on feedback.

\begin{figure}
\centering
\includegraphics[width=8cm]{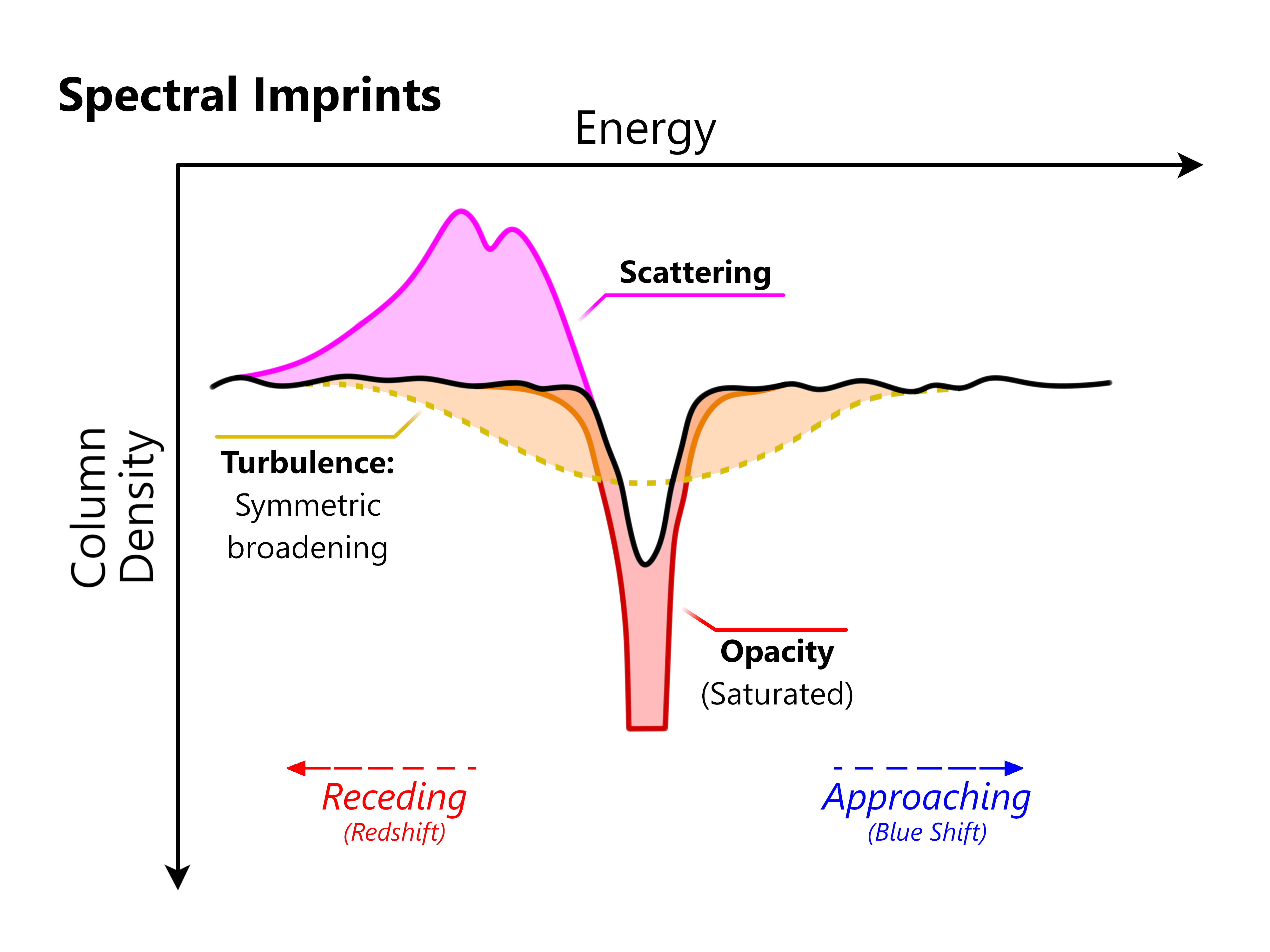}
\includegraphics[width=8cm]{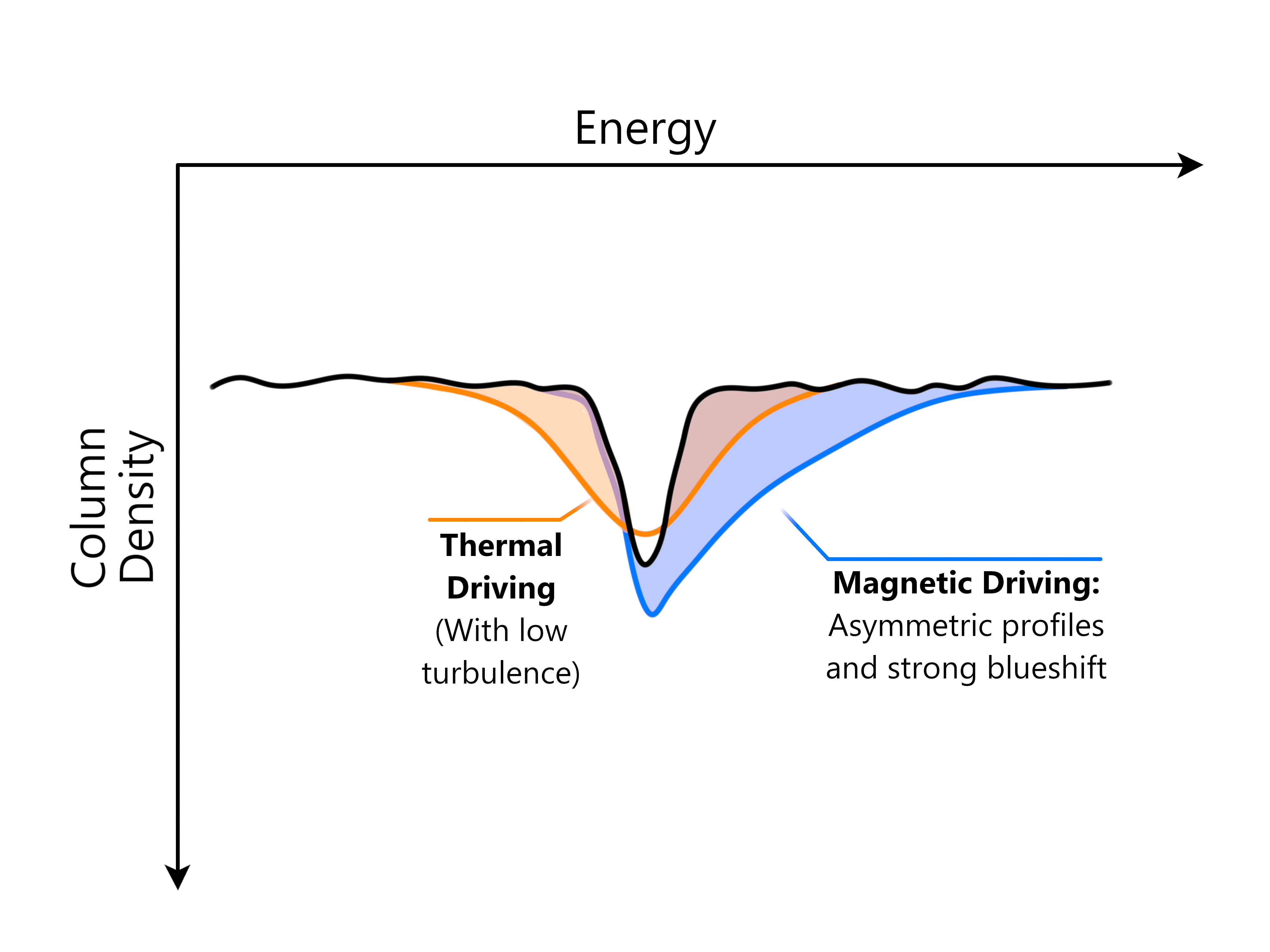}
\caption{\textcolor{teal}{Schematic illustration of the spectral line imprints left by various feedback mechanisms, gas kinematics, and the ambient physical conditions in the accretion system.} \textcolor{mypurple}{These influence the position of the line (depicted as Energy on the x-axis), its strength (depicted as a column density on the y-axis), as well as its width. 
\underline{Upper panel:} First, we show how line characteristics translate into some generic physical effects. Turbulence results in broadening while a flat bottom denotes line-of-sight optically-thick absorption. Scattering can mimic apparent redshifted emission and P\,Cygni--like profiles. Line shifts due to approaching and receding motions are indicated. \underline{Lower panel:} Next, we show profiles expected for isolated lines driven by thermal and magnetic winds. The latter are expected to preferentially show stronger blue skews (based on \cite{fukumura10apj}), though other shapes, including red skews, are possible 
depending on the detailed  photoionisation balance and the wind density at the innermost launching region, \cite{giustini12apj,fukumura19apj}. However, the magnetically-driven lines are expected to be significantly more asymmetric than in the case of thermal winds. The spectral resolution of microcalorimeters to be flown aboard \xrism\ and \athena\ is similar to the width of the line expected for thermal driving with low turbulence (or better) in typical Galactic XRBs, so these lines will easily be resolved.}
}\label{fig:lines}
\end{figure}

\vspace*{1cm}
\noindent
\textbf{3.2.  The Physics of Launching Outflows:} Accretion disc winds can be launched by (1) thermal, (2) magnetic or (3) radiation pressure. Which mechanism or mechanisms dominate(s), and whether XRBs and AGN can be unified in this regard, continues to be debated (e.g. \cite{proga02apj,diaztrigo16an,miller06nat,tomaru20mnras,mizumoto21mnras,laha21}). These mechanisms can leave telltale imprints on various absorption (and emission) lines in the X-ray spectral regime. But characterisation of the winds, and consequently of their origin, is hampered by current X-ray instrumentation, with limited capabilities to resolve line profiles, to disentangle different wind components, to measure small ($<$\,100\,km\,s$^{-1}$) line shifts, and to overcome saturation effects associated with high optical depths. Fig.\,\ref{fig:lines} attempts to strip down and capture some of these complex effects at play, in order to illustrate the expected imprints schematically. Importantly, due to these limitations, the density-velocity profile of winds, a key diagnostic of their origin \cite{tomaru20mnras}, remains to be measured. 

Thermal (Compton-heated) and magnetic winds differ in that there is a minimum disc radius at which the former can be launched. Such a  radius depends on the Compton temperature, the temperature to which the upper layers of the accretion disc are heated by X-rays from the innermost accretion regions, and this temperature is, in turn, uniquely determined by the illuminating spectrum \cite{begelman83apj}. Consequently, in \underline{thermal winds}, we expect a small range of velocities, set by the escape velocity at the launching radius (a few hundred to a few thousand km\,s$^{-1}$ at radii $\sim$\,10$^6$--10$^4$\,$R_{\rm G}$, typical of XRBs in the soft state). 

In contrast, it may be possible to launch \underline{magnetic winds} from across the entire disc, resulting in high speeds for inner disc lines and broad and asymmetric line profiles overall, reflecting the transverse motions that are unique to such winds \cite{contopoulos94apj,fukumura10apj,fukumura19apj}. This appears to hold true quite generally, despite a broad variety of predictions for the structure of magnetic winds that depends on the initial (uncertain) magnetic field configuration  \cite{fukumura10apjb,chakravorty16aa,waters18mnras}. 

\begin{figure}
\centering
\includegraphics[angle=0,width=9cm]{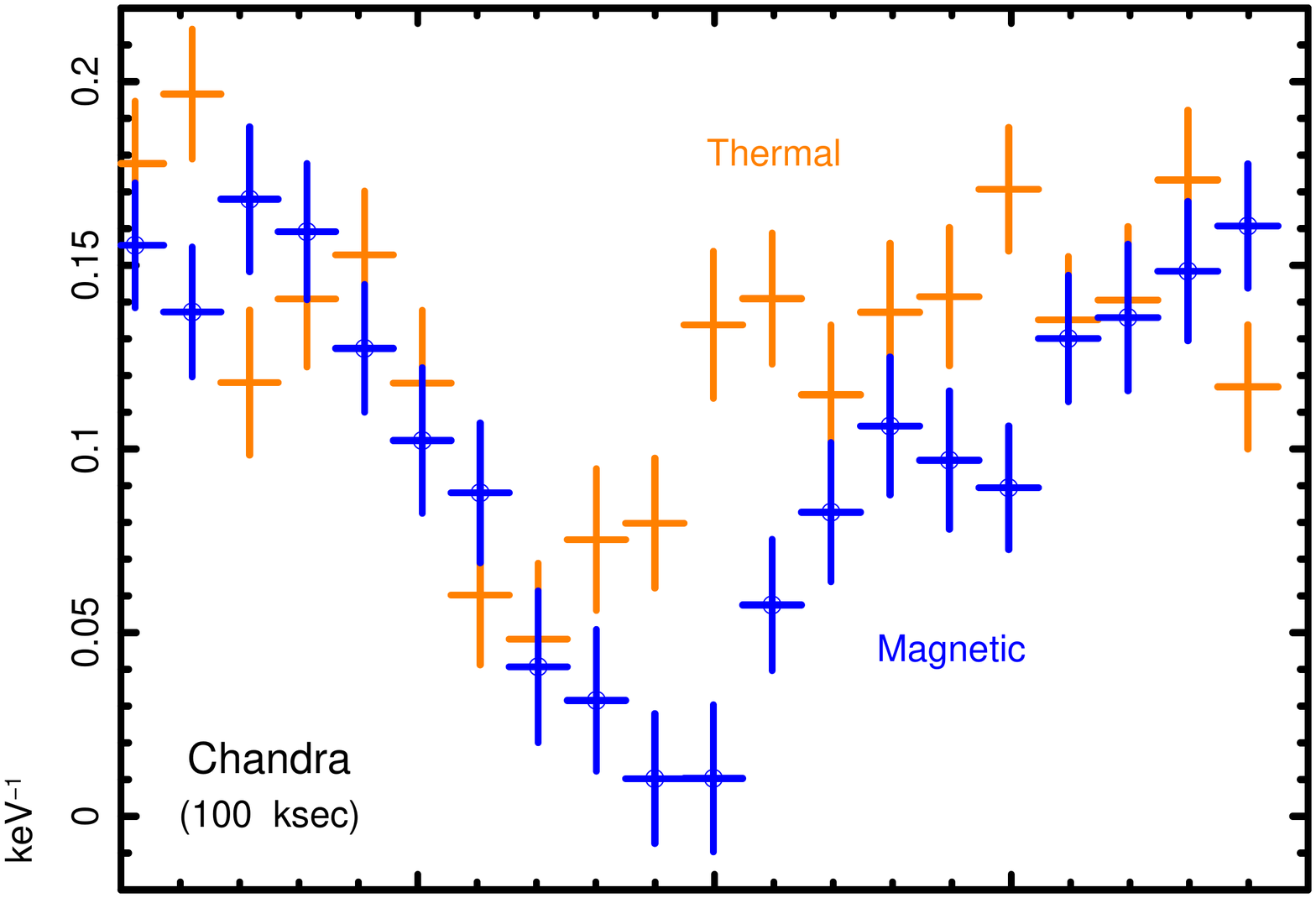}\\
\vspace*{-1.7cm}
\includegraphics[angle=0,width=9cm]{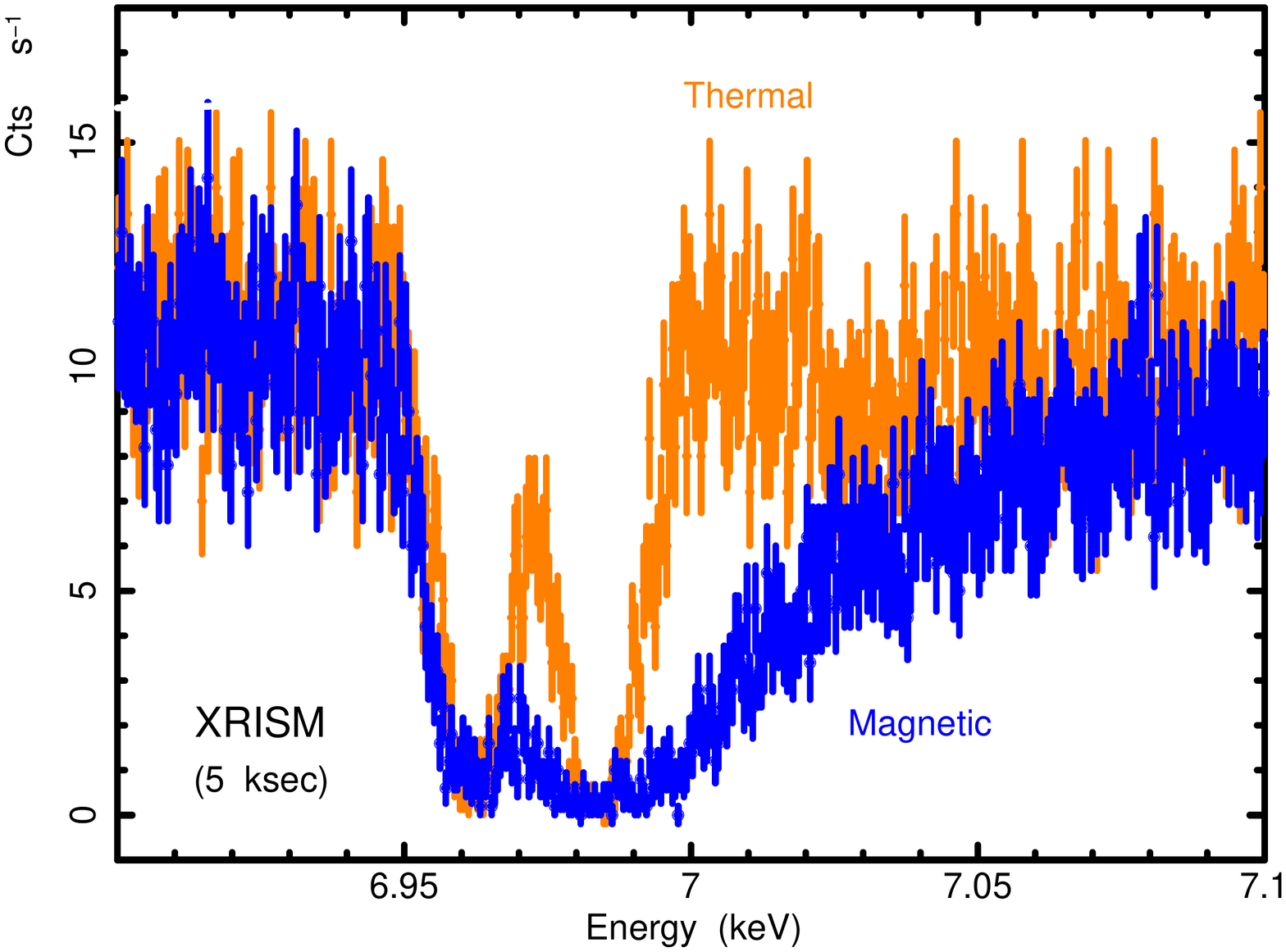}
\vspace*{0.1cm}
\caption{
\textcolor{teal}{{\em Simulations of XRB thermal vs. magnetic winds.}} \textcolor{mypurple}{
Fe\,\textsc{XXVI} doublet resulting from thermal (orange points) and magnetic (blue circled points) winds simulated with the parameters of the black hole XRB GRO\,J1655--40 during its 2005 outburst \cite{fukumura21,tomaru22}. The upper/lower panels show simulations of 100\,ksec/5\,ksec observations with the \chandra\ gratings and with the \xrism{} microcalorimeter Resolve, respectively. The \chandra\ gratings are photon-starved and cannot resolve the doublet. The asymmetry of the line profiles in the magnetic wind is also not unambiguously captured, so that the observations cannot distinguish a magnetic wind from a thermal wind with a high turbulence. \xrism\ easily overcomes these handicaps, in far shorter an exposure.
}\label{fig:windsim}}
\end{figure}

As one example of a litmus test between these competing models, Fig.\,\ref{fig:windsim} presents a simulation of \xrism's capability, relative to the current best grating spectral resolution with HETGS in the Fe band. Here, an optically thick wind has been simulated for the case of the bright 2005 outburst of the black hole XRB GRO\,J1655--40, assuming both a thermal-radiative \cite{tomaru22} and a magnetic driving model \cite{fukumura21}, respectively. The \chandra\ grating data obtained for this outburst played an influential role in triggering interest in magnetic wind driving models, but this has remained debated \cite{miller06nat, tomaru22}. The simulation zooms in on the Fe\,\textsc{xxvi} doublet at $\approx$\,6.97\,keV, arising from spin--orbit coupling. The magnetic wind extends over a wide range in radii down to regions very close in to the black hole. It thus shows the characteristic aforementioned blue asymmetry lacking in the thermal wind which originates farther out. \xrism\ will not only measure the absorbing wind properties, but will also allow the profiles to be modelled with far greater fidelity, in a fraction of the exposure time relative to the gratings \cite{tomaru22, fukumura19apj}.

In the absence of additional turbulence, the simulated line doublet can be resolved with microcalorimeters for thermal winds. This doublet offers the advantage of being free from line blends impacting other energies, but only weak constraints derived from observations of bright XRBs with third order \chandra\ gratings are currently available on this feature \cite{miller15,tomaru20mnras}. While turbulence could add additional scatter, a high degree of turbulence is not expected according to current simulations and, most importantly, isotropic turbulence will not result in asymmetric lines \cite{tomaru20mnras}  (see Fig.~\ref{fig:lines}, lower panel). Overall, this provides a promising potential pathway for distinguishing the underlying launch mechanism.

This picture can be  complicated in the presence of \underline{radiation pressure}, which may drive
winds on its own or in combination with thermal or magnetic pressure \cite{proga02apj,everett05apj,ohsuga09pasj}. Radiation pressure can manifest in multiple ways, including (i) \lq line driving\rq\ which arises from boosted absorption opacities at resonance line transitions, (ii) and \lq continuum driving\rq\ via Thompson scattering in highly ionised gas (a third scenario of \lq dusty AGN outflows\rq\ was discussed in Section\,2). In particular, line radiation pressure can drive very fast winds at relatively low luminosities, though inclusion of relativistic deboosting could reduce the expected velocities \cite{luminari20aa}. In any case, an important requirement for line driving is that the gas must have a low ionisation state in order to support the line force in the first place \cite{proga04apj}. Therefore, ionisation of the gas after it has been accelerated, or some degree of \underline{shielding} in hot discs, is required in order to explain winds that are simultaneously fast and highly ionised \cite{nomura17mnras, mizumoto21mnras, matthews20mnras}. Continuum radiation pressure is instead expected to be more efficient in launching winds at high Eddington accretion rates via Thomson scattering in fully ionised gas \cite{shakura73aa}, or to aid in launching of thermal winds closer to the compact object \cite{proga02apj}. In the end, the difficulty in disentangling the role of radiation pressure lies in the need to infer the illumination to which the gas is exposed despite complex radiative transfer effects \cite{shidatsu16apj,reeves18apj,boissay19apj,mizumoto21mnras}, and to try and recognise whether \lq failed winds\rq\ (e.g. \cite{giustini19aa, matthews20mnras}) or \lq vertically extended coronae\rq\ \cite{begelman83apjb} provide the necessary shielding for line driving in regions that would otherwise be too hot. However, even in this case, magnetic wind line profiles can be extreme (i.e. wider and more asymmetric), helping to disentangle the roles of radiation and magnetic pressure \cite{reeves18apj, boissay19apj, mizumoto21mnras}.

Proper characterisation of optically thick regions, with equivalent Hydrogen column densities $>$\,$10^{23}$--10$^{24}$\,cm$^{-2}$, is often impossible due to line saturation (see Fig.~\ref{fig:lines}, upper panel). With microcalorimeters, line saturation (and thus \underline{opacity}) will be reliably measured and lines in crowded regions of the spectrum will be resolved, allowing multiple wind components to be disentangled. Small redshifts signaling the possible presence of failed winds will be then easily identifiable. Crucially, line profiles will be properly resolved in uncrowded regions of the spectrum, such as the Fe\,\textsc{xxvi} band. Here, the {\textsl{emission}} component via \underline{scattering} in the wind also becomes relevant. This emission component will be broad and centred at the rest wavelength, thus reducing the equivalent width of the blueshifted absorption \cite{sim08mnras}. A precise characterisation of the full profile of the lines will then constrain the scattering fraction, the wind launching radius, and the solid angle, crucial for measuring the mass outflow rate. Determination of the illumination of parts of the disc/wind shielded from the central source will provide clues on the long-standing mystery of how nature prevents wind over-ionisation.

Obscuring and scattering columns in some sources are known to be exceptionally variable. Classic examples include XRBs that approach high Eddington accretion rates such as V404\,Cyg, which showed spectacular \underline{spectral variability} on a broad range of timescales during outburst \cite{motta17}. It remains unclear how much of this is intrinsic, or caused by variable gas columns and changes in ionisation degree, or a combination of these. Time-resolved high-resolution spectroscopy of key obscuration diagnostics such as the Compton shoulder and absorption edges, obtained during future bright outbursts, should help to resolve this. 

Ultimately, line profiles will likely be more complex in reality than depicted in Fig.~\ref{fig:lines} due to limitations of current simulations, e.g. in the treatment of radiation transfer for optically thick winds, or the inclusion of scattering. However, detailed simulations for different types of winds in AGN already reveal the extent to which precisely determining saturation, measuring line profiles, and resolving different wind components with microcalorimeters, together with some prior knowledge of systems such as inclination (well constrained for a majority of XRBs) will help to determine the kinematic structure of accretion disc winds for the first time, and to pinpoint their origin \cite{giustini12apj}.

\section{Concluding Remarks}
If \hitomi\ unlocked the door to X-ray microcalorimetric studies, \xrism{} is expected to throw this door open wide. The decade of the 2030s should then be ripe for detailed exploration of this new parameter space with the next generation of missions, hopefully improving upon \xrism's spectral resolution by a factor of 2 or more \cite{athena, barret18}. The studies outlined herein represent only the tip of the iceberg in terms of microcalorimeter advancements expected for accretion studies. Unforeseen and exciting serendipitous discoveries beyond these can be expected whenever new parameter space is opened up \cite{harwit81}. 

We have focused on science that only high spectral resolution X-ray studies can deliver. But accretion is an inherently broadband phenomenon, and there is no doubt that multiwavelength observations coordinated with microcalorimeters will yield novel constraints on a host of important questions, whether it be the link between winds in various ionisation stages in AGN and XRBs \cite{tombesi15, v404:munoz-darias16nature}, the distribution of dust vs. gas in AGN \cite{Esparza-Arredondo2021} or the accretion--jet connection on timescales relevant for the inner accretion flows \cite{gandhi17}, to name just a few outstanding issues. The newly operational {\em James Webb Space Telescope} is also expected to play to pivotal role in addressing these issues, enabling highly sensitive and high spatial resolution mid-infrared constraints that can be compared to X-ray observations.

But all this will only come to fruition if the instruments operate successfully in space, and if systematic uncertainties (both instrumental and modeling) can be kept under control. At the time of writing, a redefinition exercise to optimise \athena{} science under a new mission designation {\em NewAthena} is being conducted, where such considerations will be paramount. \hitomi\ data highlighted discrepancies of order 16\% between Fe abundance estimates from different atomic databases  \cite{hitomi_atomicdata}, significantly exceeding statistical errors. A putative 3.5\,keV feature in deep observations of galaxy clusters and galactic haloes may be a signature of decaying sterile neutrino dark matter particles \cite{bulbul14}. But this also remains controversial, and ill-appreciated processes such as charge exchange together with abundance mismatches could provide alternative explanations \cite{hitomi_darkmatter, gu18}. Hand-in-hand with better modeling and laboratory validation needed to resolve systematic conflicts, adequate resources to support calibration activities will be crucial to ensure optimal science exploitation \cite{mehdipour16}.


\section{Author Contributions}
The contact authors are MDT, TK and PG, all of whom  contributed equally to the main scientific content presented herein. PG led the conception, coordination and thematic definition of this work, together with participants of the XCalibur2019 international workshop at Winchester, UK, in July 2019\footnote{http://www.astro.soton.ac.uk/xcal2019/}. The focus of the workshop was to identify and review game-changing science to be expected in forthcoming microcalorimeter missions, serving as a genesis for this review. JAP executed the artistic design of the schematic diagram and contributed to the design of all other figures herein. TK is responsible for Section\,2 and MDT led Section\,3. All authors read and contributed to the overall text. 

\section{Acknowledgements}
We thank two anonymous reviewers for help in improving the content discussion. Dedicated to those who have worked to make microcalorimeter science a reality. Partially supported by the University of Southampton STAG (Theory, Astronomy and Gravity) Research Centre.

\bibliographystyle{naturemag}
\bibliography{gandhi22_natast}

\end{document}